\documentclass[%
 reprint,
 amsmath,amssymb,
 aps,
]{revtex4-2}

\usepackage[utf8]{inputenc}

\usepackage{graphicx}% Include figure files
\usepackage{dcolumn}% Align table columns on decimal point
\usepackage{bm}% bold math
\begin{document}

\preprint{APS/123-QED}

\title{Physics-guided residual correction of $\alpha$-decay half-lives based on the effective liquid drop model }

\author{Qingning Yuan}
%\email{yuan_QN@163.com}
\author{Xuanpeng Xiao}
%\email{xxp232316@163.com}
\author{Panpan Qi}
%\email{qipanpan1999@163.com}
\author{Anqi Yang}
%\email{18288226459@163.com}
\author{Gongming Yu}
  \email{ygmanan@kmu.edu.cn}
\affiliation{%
College of Physics and Technology, Kunming University, Kunming 650214, China
}%

\author{Haitao Yang}
  \email{yanghaitao205@163.com}
\affiliation{%
School of Physics and Information Engineering, Zhaotong University, Zhaotong 657000, China
}%

\author{Zhangyan Li}
 \email{20230001@ztu.edu.cn}
\affiliation{%
School of Physics and Information Engineering, Zhaotong University, Zhaotong 657000, China
}%

\author{Yanbing Cai}
  \email{yanbingcai@mail.gufe.edu.cn}
\affiliation{%
Key Laboratory of Economic System Simulation of Guizhou Province, Guizhou University of Finance and Economics, Guiyang 550025, China
}%

\begin{abstract}

To improve the prediction accuracy of $\alpha$-decay half-lives in heavy and superheavy nuclei, a physics-guided residual-correction framework combining the effective liquid drop model (ELDM) with machine-learning methods is proposed. The ELDM is first used as the macroscopic baseline for describing the barrier-penetration process, and XGBoost and TabPFN models are then employed to learn the residual deviations between ELDM predictions and experimental data. To incorporate microscopic nuclear-structure information, several physically motivated descriptors are constructed, including deformation-related quantities, Geiger--Nuttall-related features, and minimum orbital angular momentum.
The results show that machine-learning residual correction significantly improves the predictive performance of the ELDM baseline. Among all models, TabPFN-term3 achieves the best accuracy, reducing the RMSE and MAE to 0.348 and 0.248, corresponding to improvements of 38.60\% and 40.46\%, respectively. Residual-distribution and feature-ablation analyses further indicate that the corrected predictions are closer to experimental values and that physically motivated descriptors play an important role in learning nonlinear residual structures.
Overall, the proposed ELDM-based residual-correction framework can effectively compensate for missing microscopic nuclear-structure effects while preserving physical interpretability, providing a feasible strategy for high-precision $\alpha$-decay half-life prediction.

\end{abstract}

%\keywords{Suggested keywords}%Use showkeys class option if keyword
                              %display desired
\maketitle

%\tableofcontents

\section{INTRODUCTION}

$\alpha$ decay is one of the dominant radioactive decay modes of heavy and superheavy nuclei and plays an essential role in nuclear-structure studies, nuclear-stability investigations, and the identification of new nuclides~\cite{geiger1911lvii,gamow1928quantentheorie,oganessian2015super,hofmann2000discovery}. Owing to the strong sensitivity of the $\alpha$-decay half-life to the nuclear binding energy, barrier structure, and microscopic nuclear-structure effects, $\alpha$ decay provides valuable information on shell effects, pairing correlations, and nuclear deformation properties. In particular, $\alpha$ decay is widely regarded as one of the most important probes for exploring the possible ``island of stability'' in the superheavy region. With the continuous development of experimental techniques and the synthesis of new superheavy nuclei, increasing amounts of experimental $\alpha$-decay data have become available for heavy and superheavy systems~\cite{royer2000alpha,abdullin2010synthesis,poenaru2011heavy,denisov2009alpha,sobiczewski2007description}. Consequently, developing physically interpretable theoretical frameworks with improved predictive accuracy for $\alpha$-decay half-lives remains an important topic in contemporary nuclear physics.

From a theoretical perspective, $\alpha$ decay can be regarded as a quantum tunneling process in which an $\alpha$ cluster penetrates the interaction barrier and subsequently escapes from the parent nucleus. Based on this physical picture, a variety of theoretical approaches have been developed to describe $\alpha$-decay half-lives. Among them, the Geiger--Nuttall (GN) law first revealed the empirical correlation between the $\alpha$-decay half-life and decay energy~\cite{geiger1911lvii}, while the Gamow theory further established the quantum-tunneling interpretation of $\alpha$ decay~\cite{gamow1928quantentheorie}. Subsequently, many theoretical models based on the WKB approximation have been widely applied to $\alpha$-decay studies, including the universal decay law (UDL), double-folding models, density-dependent cluster models, generalized liquid drop models (GLDM), effective liquid-drop-type models, and related microscopic or semi-microscopic approaches~\cite{qi2009universal,royer2000alpha,qi2009universal,zdeb2013half,zdeb201311half,zdeb2014half,Zdeb2014ActaPhysPolon}. 
Among these approaches, liquid-drop-type models have attracted considerable attention because they provide a physically intuitive macroscopic description of the barrier-penetration process in nuclear decay. In particular, the effective liquid drop model (ELDM) describes $\alpha$ decay within a fission-like physical picture, in which the parent nucleus evolves toward a dinuclear configuration composed of the emitted $\alpha$ particle and the daughter nucleus, and the corresponding barrier penetrability is evaluated through the semiclassical WKB approximation. Owing to its clear physical interpretation and its capability to describe nuclear decay half-lives within a unified macroscopic framework, the ELDM provides a suitable physical baseline for constructing a machine-learning residual-correction model in the present work~\cite{zhang2006alpha,bao2014systematical,ni2008unified,goncalves1993effective,gonccalves1997prescold,tavares1998effective,pomorski2015spontaneous,zdeb2016proton,Pomorski2017NuclTheory}.

Despite the clear physical picture and broad applicability of the ELDM in describing nuclear decay half-lives, conventional ELDM calculations still exhibit several limitations. As a macroscopic liquid-drop-type model, the ELDM mainly describes the global evolution of the decay system and the corresponding barrier-penetration behavior, while microscopic nuclear-structure effects such as deformation, shell effects, pairing correlations, and angular-momentum hindrance are not always explicitly incorporated in a fully microscopic manner~\cite{royer2000alpha,zhang2006alpha,pahlavani2013effect,akrawy2019influence}. In addition, quantities related to the formation probability and microscopic configuration of the emitted $\alpha$ cluster are usually treated in an effective or simplified way in practical calculations. Consequently, although the ELDM can reasonably reproduce the overall tunneling trend of $\alpha$ decay, non-negligible residual deviations may still remain for many nuclei, especially for odd-$A$ and odd-odd systems~\cite{qi2009universal}.

In recent years, machine-learning methods have attracted increasing attention in nuclear physics research and have gradually been applied to nuclear mass prediction, nuclear radius estimation, decay-property analysis, and nuclear reaction studies~\cite{boehnlein2022colloquium,niu2022nuclear,gao2021machine}. Compared with traditional empirical formulas and parameterized theoretical models, machine-learning approaches possess a stronger capability for extracting complex nonlinear correlations hidden in high-dimensional nuclear datasets, thereby demonstrating significant potential for improving the predictive accuracy of nuclear properties.

Nevertheless, many existing machine-learning studies on $\alpha$ decay mainly rely on purely data-driven fitting strategies in which the half-lives are directly predicted without sufficiently incorporating the underlying physical mechanisms governing the decay process. Although such approaches can improve predictive accuracy to some extent, their physical interpretability often remains limited. In particular, the physical origins of the residual deviations, as well as the specific roles played by microscopic nuclear-structure effects in the prediction process, are still not fully understood. Therefore, developing machine-learning frameworks that can simultaneously maintain predictive accuracy and physical interpretability remains an important challenge in current $\alpha$-decay studies~\cite{gao2023studies,qu2025nuclear,he2023machine}.

To address the above issues, a physics-guided residual-learning framework based on the effective liquid drop model is developed in the present work. First, the ELDM is employed as the baseline description of the macroscopic decay process, and the corresponding theoretical half-lives are calculated accordingly. Subsequently, the discrepancies between the ELDM predictions and experimental data are defined as residual quantities and further learned through machine-learning approaches~\cite{karniadakis2021physics,cuomo2022scientific}. Unlike purely black-box strategies that directly fit $\alpha$-decay half-lives, the present framework mainly focuses on the missing microscopic structure effects that are not explicitly incorporated in the baseline ELDM calculations. In this way, the machine-learning stage is primarily responsible for correcting the residual deviations originating from microscopic nuclear-structure effects, while the global decay trend remains constrained by the underlying physical model. Therefore, the present approach can effectively improve the predictive accuracy of $\alpha$-decay half-lives while preserving the physical interpretability of the overall framework~\cite{boehnlein2022colloquium,hollmann2025accurate}.

To further investigate the influences of different physical descriptors on the residual-correction process, several groups of physically motivated input features are systematically constructed in the present work, including nucleon numbers, decay energy, deformation-related quantities, Coulomb--decay-energy competition terms, and minimum orbital angular momentum. These descriptors are designed to characterize both the global decay behavior and the microscopic nuclear-structure effects that are not explicitly incorporated in the baseline ELDM calculations. In particular, deformation- and angular-momentum-related features are introduced to describe the missing microscopic corrections associated with barrier-structure modification and centrifugal hindrance effects in unfavored $\alpha$ decay. Furthermore, two representative small-sample machine-learning models, namely XGBoost and TabPFN, are systematically compared for $\alpha$-decay residual prediction in order to investigate their capabilities in learning complex nonlinear correlations hidden in nuclear-structure information. Through the comparison of different feature combinations and learning frameworks, the present work aims to further clarify the roles of physically motivated descriptors in the residual-learning correction of $\alpha$-decay half-lives~\cite{hollmann2025accurate,pahlavani2013effect,akrawy2019influence}.

The remainder of this paper is organized as follows. Section II introduces the ELDM framework, the residual-learning methodology, and the physically motivated feature-construction strategy adopted in the present work. Section III presents the dataset, computational details, and the corresponding numerical results of the machine-learning residual corrections, including the analyses of residual distributions and feature-ablation behaviors. Finally, the main conclusions of the present study are summarized in Section IV.

\section{GENERAL FORMALISM}
\subsection{ELDM framework}

In the present work, the effective liquid drop model (ELDM) is adopted as the baseline physical framework for describing $\alpha$-decay half-lives. In the ELDM picture, $\alpha$ decay is treated as a highly asymmetric fission-like process rather than as a purely external motion of a preformed $\alpha$ particle. During the decay process, the parent nucleus gradually evolves toward a dinuclear configuration composed of a light fragment and a heavy fragment. For $\alpha$ decay, the light fragment corresponds to the emitted $\alpha$ particle, whereas the heavy fragment corresponds to the daughter nucleus~\cite{goncalves1993effective,gonccalves1997prescold,tavares1998effective}.

\begin{figure*}[ht]
\centering
\includegraphics[width=0.5\textwidth]{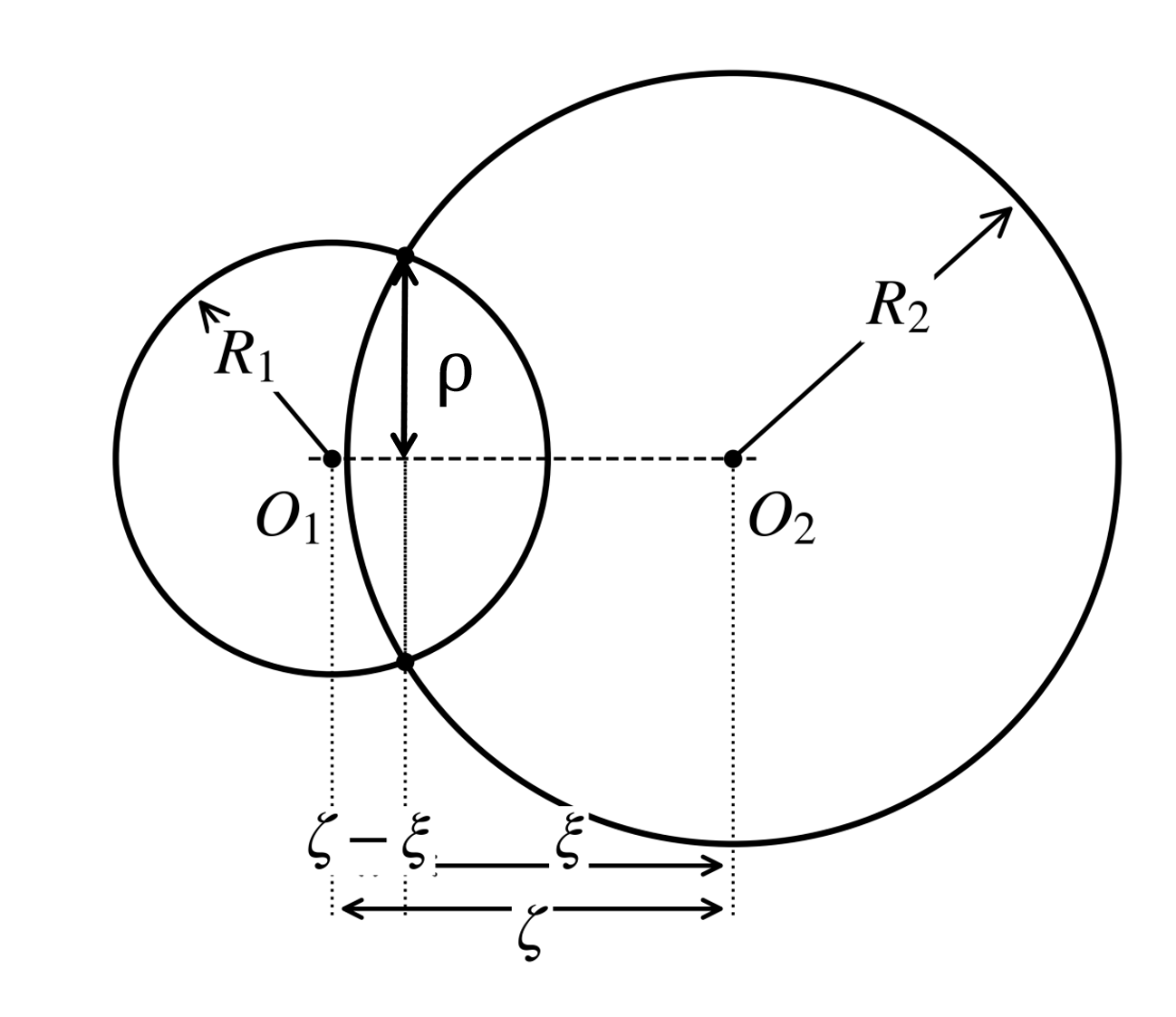}
\caption{\label{fig:eldm_geometry}
Schematic illustration of the prescission dinuclear configuration adopted in the effective liquid drop model. The two intersecting spherical fragments are characterized by the radii \(R_1\) and \(R_2\), the center-to-center distance \(\zeta\), and the distance \(\xi\) from the intersection plane to the center of the heavier fragment.}
\end{figure*}

In the prescission region, the dinuclear system is described by two intersecting spherical fragments~\cite{duarte1996effective,sheng2011systematic}. The corresponding geometrical configuration is characterized by four shape parameters,
\begin{equation}
\{R_1,R_2,\zeta,\xi\},
\end{equation}
where $R_1$ and $R_2$ denote the radii of the light and heavy spherical fragments, respectively. The quantity $\zeta$ represents the distance between the geometrical centers of the two spherical fragments, while $\xi$ denotes the distance from the intersection plane to the geometrical center of the heavier fragment. Accordingly, the distance from the intersection plane to the geometrical center of the lighter fragment is $\zeta-\xi$. A schematic illustration of this two-sphere configuration is shown in Fig.~\ref{fig:eldm_geometry}.

The four geometrical parameters are not independent. If all of them were allowed to vary freely, the deformation path of the dinuclear system would not be uniquely determined. Therefore, several constraint relations are introduced in the ELDM to reduce the multidimensional shape evolution to an effective one-dimensional deformation path along the collective coordinate $\zeta$. For a given value of $\zeta$, the remaining geometrical quantities can be determined through the corresponding constraint equations.

The first constraint is the volume-conservation condition. Since the nuclear liquid is approximately incompressible, the total volume of the dinuclear configuration should remain equal to that of the original parent nucleus during the deformation process~\cite{goncalves1993effective,sheng2015investigation,tavares1998effective}. This condition can be written as
\begin{equation}
2(R_1^3+R_2^3)
+3\left[R_1^2(\zeta-\xi)+R_2^2\xi\right]
-\left[(\zeta-\xi)^3+\xi^3\right]
-4R_m^3=0,
\label{eq:eldm_volume_constraint}
\end{equation}
where $R_m$ denotes the radius of the parent nucleus. This relation is obtained by adding the volumes of the two spherical segments separated by the intersection plane and requiring their sum to be equal to the volume of the parent nucleus.

More specifically, the volume of the light-fragment spherical segment can be expressed as
\begin{equation}
V_1=
\frac{\pi}{3}
\left[
2R_1^3
+3R_1^2(\zeta-\xi)
-(\zeta-\xi)^3
\right],
\end{equation}
and the volume of the heavy-fragment spherical segment is
\begin{equation}
V_2=
\frac{\pi}{3}
\left[
2R_2^3
+3R_2^2\xi
-\xi^3
\right].
\end{equation}
The volume of the original parent nucleus is
\begin{equation}
V_m=\frac{4}{3}\pi R_m^3.
\end{equation}
By imposing
\begin{equation}
V_1+V_2=V_m,
\end{equation}
and eliminating the common factor $\pi/3$, Eq.~\eqref{eq:eldm_volume_constraint} is obtained. Thus, the first constraint ensures that the nuclear matter volume is conserved during the prescission deformation.

The second constraint is the geometrical matching condition at the intersection plane. In three-dimensional space, two intersecting spheres share the same circular intersection window. Therefore, the radius of this intersection circle must be identical whether it is calculated from the light-fragment sphere or from the heavy-fragment sphere. If the radius of the intersection circle is denoted by $\rho$, then
\begin{equation}
\rho^2=R_1^2-(\zeta-\xi)^2
\end{equation}
for the light-fragment sphere, and
\begin{equation}
\rho^2=R_2^2-\xi^2
\end{equation}
for the heavy-fragment sphere~\cite{goncalves1993effective,sheng2015investigation,tavares1998effective}. Since these two expressions correspond to the same circular window, one obtains
\begin{equation}
R_1^2-(\zeta-\xi)^2
=
R_2^2-\xi^2.
\end{equation}
This relation can be rewritten as
\begin{equation}
R_1^2-R_2^2-(\zeta-\xi)^2+\xi^2=0.
\label{eq:eldm_geometry_constraint}
\end{equation}
Equation~\eqref{eq:eldm_geometry_constraint} guarantees that the two spherical fragments are geometrically matched at the same intersection circle.

The third constraint specifies the mass-asymmetry evolution mode of the dinuclear system. In the varying-mass-asymmetry shape (VMAS) prescription, the radius of the light-fragment generating sphere is fixed at its final-fragment value during the prescission evolution~\cite{goncalves1993effective,sheng2015investigation,tavares1998effective}. This condition is expressed as
\begin{equation}
R_1-\overline{R}_1=0,
\label{eq:eldm_vmas_constraint}
\end{equation}
or equivalently,
\begin{equation}
R_1=\overline{R}_1,
\end{equation}
where $\overline{R}_1$ denotes the radius of the light fragment in the final separated configuration. For $\alpha$ decay, this light fragment corresponds to the emitted $\alpha$ particle. It should be emphasized that Eq.~\eqref{eq:eldm_vmas_constraint} fixes the radius of the generating sphere associated with the light fragment, rather than implying that a completely separated $\alpha$ particle already exists at every stage of the prescission evolution.

Through the above three constraints, the original four geometrical variables \(R_1\), \(R_2\), \(\zeta\), and \(\xi\) are reduced to an effective one-dimensional deformation path along \(\zeta\). For each value of \(\zeta\), the corresponding dinuclear configuration can be determined by solving the constraint equations. The effective potential barrier and the inertia coefficient are then constructed along this deformation path for the subsequent Gamow/WKB penetrability calculation.

After the geometrical constraints are imposed, the ELDM reduces the shape evolution of the parent nucleus toward the $\alpha$-particle--daughter configuration to a one-dimensional barrier-penetration problem along the collective coordinate $\zeta$. Along this path, the potential barrier is described in two regions: the inner liquid-drop barrier in the prescission region and the external Coulomb barrier after the touching configuration is reached.

In the prescission region, the two fragments are still described by an intersecting two-sphere configuration. The inner barrier is defined as the energy difference between the current dinuclear configuration and the final separated-fragment configuration. It is given by the sum of the Coulomb-energy change and the surface-energy change~\cite{duarte1998cold,gonccalves2017two,ashok2024half}:
\begin{equation}
V_{\rm in}(\zeta)
=
\left[E_C(\zeta)-E_C^{(12)}\right]
+
\left[E_S(\zeta)-E_S^{(12)}\right],
\label{eq:vin_eldm}
\end{equation}
where $E_C(\zeta)$ and $E_S(\zeta)$ are the Coulomb and surface energies of the intersecting two-sphere configuration at a given $\zeta$, respectively. The quantities $E_C^{(12)}$ and $E_S^{(12)}$ denote the sums of the Coulomb self-energies and surface energies of the two final separated fragments. Therefore, the inner barrier represents the change in the macroscopic liquid-drop energy during the prescission evolution.

The effective surface-tension coefficient $\sigma_{\rm eff}$ entering the surface-energy term is constrained by the decay energy $Q_\alpha$ and can be written as~\cite{zhang2006theoretical,wang2017competition}:
\begin{equation}
\sigma_{\rm eff}
=
\frac{
Q_\alpha+E_C^{(12)}-E_C^{(m)}
}{
4\pi\left(R_p^2-R_1^2-R_2^2\right)
},
\label{eq:sigma_eff}
\end{equation}
where $E_C^{(m)}$ is the Coulomb self-energy of the spherical parent nucleus, $R_p$ is the parent-nucleus radius, and $R_1$ and $R_2$ are the radii of the light and heavy fragments, respectively. Through this effective treatment, the experimental decay energy is incorporated into the macroscopic surface-energy contribution.

After the touching configuration is reached,
\begin{equation}
R_{\rm touch}=R_1+R_2,
\label{eq:rtouch}
\end{equation}
the system enters the external separated region. In this region, the emitted $\alpha$ particle and the daughter nucleus are treated as two separated charged bodies, and the external Coulomb barrier is expressed as~\cite{dong2011new}:
\begin{equation}
V_C^{\rm out}(\zeta)
=
\frac{Z_1Z_2e^2}{\zeta}.
\label{eq:vout_coulomb}
\end{equation}
For $\alpha$ decay,
\begin{equation}
Z_1=Z_\alpha=2,\qquad Z_2=Z_d=Z_p-2,
\end{equation}
and thus
\begin{equation}
V_C^{\rm out}(\zeta)
=
\frac{2Z_de^2}{\zeta}.
\label{eq:vout_alpha}
\end{equation}

For transitions with nonzero orbital angular momentum, the centrifugal contribution is further included as~\cite{liu2025effect,choi2025correlation}:
\begin{equation}
V_l(\zeta)
=
\frac{(\hbar c)^2l(l+1)}
{2\mu\zeta^2},
\label{eq:centrifugal_eldm}
\end{equation}
where $\mu$ is the two-body reduced mass of the emitted $\alpha$ particle and the daughter nucleus,
\begin{equation}
\mu=\frac{m_1m_2}{m_1+m_2}.
\label{eq:reduced_mass}
\end{equation}
Consequently, the external effective barrier can be written as
\begin{equation}
V_{\rm ext}(\zeta)
=
\frac{Z_1Z_2e^2}{\zeta}
+
\frac{(\hbar c)^2l(l+1)}
{2\mu\zeta^2}.
\label{eq:vext_eldm}
\end{equation}
For favored $\alpha$-decay transitions with $l=0$, the centrifugal contribution vanishes and the external barrier reduces to the pure Coulomb form,
\begin{equation}
V_{\rm ext}(\zeta)
=
\frac{Z_1Z_2e^2}{\zeta}.
\end{equation}

Thus, the ELDM barrier consists of the inner liquid-drop barrier $V_{\rm in}(\zeta)$ in the prescission region and the external barrier $V_{\rm ext}(\zeta)$ in the separated region. These barriers are subsequently used in the WKB/Gamow action integral to calculate the penetrability and the corresponding $\alpha$-decay half-life.

After the effective one-dimensional barrier is constructed, the barrier
penetrability is evaluated within the Gamow/WKB approximation. For a given
$\alpha$-decay energy \(Q_\alpha\), the penetrability is written as~\cite{gurney1929quantum,gurney1928wave,buck1990new}:
\begin{equation}
P=
\exp\left\{
-\frac{2}{\hbar}
\int_{\zeta_0}^{\zeta_C}
\sqrt{
2\mu_{\rm eff}(\zeta)
\left[
V(\zeta)-Q_\alpha
\right]
}
\,d\zeta
\right\}.
\label{eq:penetrability}
\end{equation}
Here, \(V(\zeta)\) denotes the effective ELDM barrier along the collective
coordinate \(\zeta\), and \(\mu_{\rm eff}(\zeta)\) is the effective inertial
coefficient associated with the collective motion along the one-dimensional
deformation path~\cite{duarte1996effective}. The quantities \(\zeta_0\) and \(\zeta_C\) represent the
inner and outer classical turning points, respectively.

In the VMAS parametrization, the inner turning point is determined by the
two-sphere geometry as~\cite{kelkar2007critical}:
\begin{equation}
\zeta_0=R_p-\overline{R}_1,
\label{eq:zeta0}
\end{equation}
where \(R_p\) is the parent-nucleus radius and \(\overline{R}_1\) is the
radius of the final light fragment. For \(\alpha\) decay, the light fragment
corresponds to the emitted \(\alpha\) particle, and thus
\begin{equation}
\overline{R}_1=R_\alpha .
\end{equation}

The outer turning point is determined by the condition
\begin{equation}
V_{\rm ext}(\zeta_C)=Q_\alpha .
\end{equation}
In the external separated region, the effective barrier consists of the
Coulomb and centrifugal contributions,
\begin{equation}
V_{\rm ext}(\zeta)
=
\frac{Z_\alpha Z_de^2}{\zeta}
+
\frac{l(l+1)(\hbar c)^2}{2\mu\zeta^2},
\end{equation}
where \(Z_\alpha=2\), \(Z_d=Z_p-2\), \(l\) is the orbital angular momentum
carried by the emitted \(\alpha\) particle, and \(\mu\) is the two-body
reduced mass of the \(\alpha\)-daughter system. Solving
\(V_{\rm ext}(\zeta_C)=Q_\alpha\) gives~\cite{denisov2005alpha,mohr2006alpha}:
\begin{equation}
\zeta_C
=
\frac{Z_\alpha Z_d e^2}{2Q_\alpha}
+
\sqrt{
\left(
\frac{Z_\alpha Z_d e^2}{2Q_\alpha}
\right)^2
+
\frac{l(l+1)(\hbar c)^2}{2\mu Q_\alpha}
}.
\label{eq:zetac}
\end{equation}

For favored \(\alpha\)-decay transitions with \(l=0\), the centrifugal
contribution vanishes, and the outer turning point reduces to
\begin{equation}
\zeta_C=
\frac{Z_\alpha Z_de^2}{Q_\alpha}.
\end{equation}
Therefore, the penetrability in the ELDM is determined by the WKB action
integral between \(\zeta_0\) and \(\zeta_C\), where the classically forbidden
region satisfies \(V(\zeta)>Q_\alpha\).

After the barrier penetrability \(P\) is obtained, the decay constant is written as
\begin{equation}
\lambda=\lambda_0 P,
\label{eq:decay_constant}
\end{equation}
where \(\lambda_0\) is the pre-exponential factor characterizing the effective assault frequency on the barrier. In the present calculation, it is taken as~\cite{zhang2011assault}:
\begin{equation}
\lambda_0=1.0\times10^{22}\ {\rm s}^{-1}.
\label{eq:lambda0}
\end{equation}
The corresponding \(\alpha\)-decay half-life is then given by
\begin{equation}
T_{1/2}
=
\frac{\ln 2}{\lambda}.
\label{eq:half_life_basic}
\end{equation}
Substituting Eq.~\eqref{eq:decay_constant} into Eq.~\eqref{eq:half_life_basic}, one obtains~\cite{nguyen2026global,royer2008recent}:
\begin{equation}
T_{1/2}
=
\frac{\ln 2}{\lambda_0P}.
\label{eq:half_life}
\end{equation}

Since \(\alpha\)-decay half-lives usually span many orders of magnitude, the calculated half-life is expressed in logarithmic form for comparison with experimental data:
\begin{equation}
\log_{10}T_{1/2}^{\rm ELDM}
=
\log_{10}
\left(
\frac{\ln 2}{\lambda_0}
\right)
-
\log_{10}P.
\label{eq:log_half_life}
\end{equation}
Therefore, the ELDM-predicted half-life is directly related to the barrier penetrability obtained from the WKB/Gamow calculation.

\subsection{Residual-learning framework}

Although the ELDM can reasonably reproduce the main barrier-penetration behavior and the global trend of $\alpha$-decay half-lives, systematic deviations between theoretical predictions and experimental data still remain. These deviations mainly arise from the macroscopic nature of the ELDM, in which microscopic nuclear-structure effects can only be partially incorporated in an effective manner through the model parameters and barrier construction. In particular, nuclear deformation effects, shell structures, pairing correlations, and residual angular-momentum hindrance may not be fully described by the effective liquid-drop barrier alone.

Therefore, in the present work, the ELDM calculation is regarded as the baseline description of the macroscopic barrier-penetration process, while the machine-learning stage is introduced to learn the residual corrections between the ELDM predictions and experimental half-lives. This residual-learning strategy is expected to compensate for the missing microscopic nuclear-structure information that is not fully captured by the ELDM.

To this end, the discrepancy between the ELDM prediction and the experimental half-life is defined as
\begin{equation}
\Delta
=
\log_{10}T_{\rm ELDM}
-
\log_{10}T_{\rm exp},
\end{equation}
where \(\log_{10}T_{\rm ELDM}\) and \(\log_{10}T_{\rm exp}\) denote the ELDM-calculated and experimental logarithmic half-lives, respectively. The residual \(\Delta\) is then modeled using machine-learning approaches.

After the machine-learning-predicted residual \(\Delta_{\rm ML}\) is obtained, the residual-corrected $\alpha$-decay half-life can be written as
\begin{equation}
\log_{10}T_{\rm corr}
=
\log_{10}T_{\rm ELDM}
-
\Delta_{\rm ML},
\end{equation}
where \(\log_{10}T_{\rm corr}\) represents the final residual-corrected prediction. Compared with purely data-driven fitting strategies, the present physics-guided residual-learning framework preserves the macroscopic barrier-penetration information provided by the ELDM while compensating for the missing microscopic nuclear-structure contributions, thereby improving the predictive accuracy of $\alpha$-decay half-lives.

\subsection{Feature engineering}

To further investigate the influences of different physical factors on the ELDM residual deviations, several groups of physically motivated input descriptors are systematically constructed in the present work. The basic descriptors include the proton number $Z$, neutron number $N$, and $\alpha$-decay energy $Q_{\alpha}$, which characterize the nuclear composition and the dominant driving force governing the barrier penetrability in $\alpha$ decay. Based on these quantities, deformation-related descriptors, Coulomb--decay-energy competition terms, and minimum orbital angular momentum are further introduced to characterize the microscopic nuclear-structure effects that are not explicitly incorporated in the baseline ELDM barrier framework.

Among these descriptors, the deformation-related term is introduced to reflect the influence of nuclear deformation on the barrier structure and tunneling behavior~\cite{pahlavani2013effect}, while the quantity $Z/\sqrt{Q_{\alpha}}$ is closely associated with the barrier-penetration behavior implied by the Geiger--Nuttall relation~\cite{geiger1911lvii,seif2019additional,sheline1991alpha,delion2007alpha}. In addition, the minimum orbital angular momentum $l_{min}$ is employed to characterize the centrifugal hindrance effect in unfavored $\alpha$ decay~\cite{akrawy2019influence}. By progressively incorporating different physically motivated descriptors, several residual-correction feature combinations are constructed to systematically analyze the contributions of various microscopic nuclear-structure effects to the residual-learning process.

In the present work, the nuclear deformation effect is characterized through a deformation-related descriptor defined as
\begin{equation}
x_{def}=\sqrt{\kappa_2\beta_2},
\end{equation}
where $\beta_2$ denotes the quadrupole deformation parameter~\cite{stone2016table}. To distinguish different nuclear-shape characteristics, a sign-related factor $\kappa_2$ is introduced. Specifically, $\kappa_2=+2$ is adopted for prolate nuclei, $\kappa_2=-1$ for oblate nuclei, and $\kappa_2=0$ for nearly spherical nuclei. Through this definition, the deformation-related information can be incorporated in a unified manner to reflect the influences of different nuclear shapes on the $\alpha$-decay barrier structure and the corresponding tunneling behavior~\cite{pahlavani2013effect,peltonen2008alpha}. In addition, the deformation effect is closely related to the microscopic structure evolution and angular-momentum hindrance behavior in $\alpha$ decay~\cite{akrawy2019influence,stone2016table,ma2022high,butler2023studies,garrett2022experimental,qian2012shape,manhas2005proximity}.

Based on the above physically motivated descriptors, four different groups of residual-correction feature combinations are further constructed to systematically investigate the influences of various microscopic nuclear-structure effects on the residual-correction performance. The detailed compositions of these feature sets are summarized in Table~\ref{tab:features}.

\begin{table}[htbp]
\caption{\label{tab:features}Different residual-correction feature combinations adopted in the present work.}
\begin{ruledtabular}
\begin{tabular}{lc}
Feature set & Input descriptors \\
\colrule
term1 & $\{Z,N,Q_{\alpha},x_{def}\}$ \\
term2 & $\{Z,N,Q_{\alpha},x_{def},Z/\sqrt{Q_{\alpha}}\}$ \\
term3 & $\{Z,N,Q_{\alpha},x_{def},Z/\sqrt{Q_{\alpha}},l_{min}\}$ \\
term4 & $\{Z,N,Q_{\alpha},Z/\sqrt{Q_{\alpha}},l_{min}\}$ \\
\end{tabular}
\end{ruledtabular}
\end{table}

\subsection{Machine-learning models}

In the machine-learning stage, XGBoost and TabPFN are employed to model the ELDM residual deviations~\cite{chen2016xgboost,hollmanntabpfn}. XGBoost is an ensemble-learning algorithm based on gradient-boosted decision trees and possesses strong nonlinear fitting capability together with good generalization performance for structured tabular data. In contrast, TabPFN is a probabilistic foundation model specifically designed for small tabular datasets and is capable of learning complex nonlinear correlations under limited-data conditions. Considering the relatively limited amount of available experimental $\alpha$-decay data, the predictive performances of these two representative small-sample learning models under different feature combinations are systematically compared in order to further investigate their capabilities in learning the residual deviations associated with microscopic nuclear-structure effects.

Overall, the framework developed in the present work combines the macroscopic barrier-penetration mechanism described by the ELDM approach with the nonlinear residual-correction capability of machine-learning models. Within this framework, the ELDM calculations provide the basic physical constraints governing the $\alpha$-decay process, while the machine-learning stage further learns the residual deviations originating from microscopic nuclear-structure effects that are not explicitly incorporated in the baseline calculations. Through this physics-guided residual-correction strategy, the predictive accuracy of $\alpha$-decay half-lives can be significantly improved while preserving the physical interpretability of the overall framework.

\subsection{Computational details}

For the XGBoost calculations, the residual-correction task was carried out within a regression framework based on gradient-boosted decision trees~\cite{chen2016xgboost,zhang2021study,zuhlke2025tcr}. In the present work, the residual quantity was defined as
\begin{equation}
\Delta=\log_{10}T_{\rm ELDM}-\log_{10}T_{\rm exp},
\end{equation}
and the machine-learning model was trained to predict this residual correction.

To reduce statistical fluctuations arising from random dataset partitioning, a 10-fold cross-validation strategy was adopted throughout the XGBoost calculations. The dataset was randomly shuffled using a fixed random seed of 42 to ensure the reproducibility of the calculations. The detailed hyperparameter settings adopted in the XGBoost calculations are summarized in Table~\ref{tab:xgb_params}. The predictive performance of the model was evaluated using the root-mean-square error (RMSE), mean absolute error (MAE), and coefficient of determination ($R^2$). In addition, permutation-based feature-importance analysis was further employed to investigate the relative contributions of different physically motivated descriptors to the residual-correction behavior.

\begin{table}[htbp]
\caption{\label{tab:xgb_params}Hyperparameter settings adopted in the XGBoost calculations.}
\begin{ruledtabular}
\begin{tabular}{lc}
Parameter & Value \\
\colrule
n\_estimators & 300 \\
max\_depth & 4 \\
learning\_rate & 0.05 \\
subsample & 0.85 \\
colsample\_bytree & 0.85 \\
reg\_alpha & 0.0 \\
reg\_lambda & 1.0 \\
random\_state & 42 \\
n\_splits & 10 \\
\end{tabular}
\end{ruledtabular}
\end{table}

For the TabPFN calculations, the residual-correction task was performed using the TabPFNRegressor framework~\cite{hollmann2025accurate}. Similar to the XGBoost calculations, the target quantity was defined as the residual deviation between the ELDM predictions and the experimental half-lives. A 10-fold cross-validation strategy was likewise adopted to ensure a fair comparison between different machine-learning approaches.

In the present work, the TabPFN model was directly trained on the residual-correction dataset without additional feature standardization. The random seed was fixed to 50 to maintain reproducibility during the cross-validation procedure. Owing to the relatively limited size of the present dataset, all TabPFN calculations were performed on CPU devices. Similar to the XGBoost calculations, the predictive performance of the TabPFN model was evaluated using RMSE, MAE, and $R^2$. Furthermore, permutation-based feature-importance analysis was also employed to investigate the relative contributions of different physically motivated descriptors in the residual-correction process.

Overall, the framework developed in the present work combines the macroscopic barrier-penetration mechanism described by the ELDM with the nonlinear residual-correction capability of machine-learning models. Within this framework, the ELDM calculations provide the basic physical constraints governing the $\alpha$-decay process, while the machine-learning stage further learns the residual deviations originating from microscopic nuclear-structure effects that are not fully incorporated in the baseline calculations. Through this physics-guided residual-correction strategy, the predictive accuracy of $\alpha$-decay half-lives can be improved while preserving the physical interpretability of the overall framework.

\section{Numerical results}

The experimental $\alpha$-decay data employed in the present work were mainly collected from the NUBASE2020 nuclear-property evaluation~\cite{kondev2021nubase2020}. The corresponding nuclear mass and decay-energy information were obtained from the AME2020 atomic mass evaluation~\cite{huang2021ame,12wng2021me}. In addition, the nuclear deformation parameters used in the feature-construction procedure were taken from the finite-range droplet model FRDM(2012) database~\cite{moller2016nuclear}.

After excluding nuclei with incomplete physical descriptors or unavailable experimental information, a systematic dataset containing experimentally known $\alpha$-decay nuclei was constructed for the subsequent residual-correction calculations. The resulting dataset covers a broad region of heavy and superheavy nuclei and includes both even-even and odd-mass systems, thereby providing sufficient diversity for investigating the influence of different microscopic nuclear-structure effects on the residual-correction behavior.

Based on the above databases, the final dataset includes the proton number $Z$, neutron number $N$, $\alpha$-decay energy $Q_{\alpha}$, deformation-related descriptors, minimum orbital angular momentum, experimental half-lives, and the corresponding ELDM theoretical predictions. To ensure the reliability and consistency of the machine-learning analysis, only nuclei with complete and available physical information were retained in the final calculations.

To evaluate the capability of the ELDM baseline in describing $\alpha$-decay half-lives, a systematic comparison between the theoretical predictions and experimental data was first carried out. Overall, the ELDM calculations can reasonably reproduce the global variation trend of $\alpha$-decay half-lives with decay energy, indicating that the macroscopic barrier-penetration mechanism provides an effective description of the dominant tunneling behavior in the $\alpha$-decay process.

Nevertheless, noticeable residual deviations can still be observed for a number of nuclei. These discrepancies are generally associated with microscopic nuclear-structure effects, including deformation effects, shell structures, pairing correlations, and orbital-angular-momentum hindrance mechanisms, which are not fully incorporated in the baseline ELDM calculations. Therefore, although the ELDM can capture the overall macroscopic decay behavior, relying solely on the conventional effective liquid-drop barrier calculation remains insufficient for achieving high-precision predictions of $\alpha$-decay half-lives. This further motivates the introduction of the subsequent machine-learning residual-correction framework.

To further evaluate the improvement introduced by the machine-learning residual correction, XGBoost and TabPFN models were employed to learn the ELDM residual deviations under different physically motivated feature combinations. The corresponding RMSE and MAE values of the residual-corrected $\alpha$-decay half-life predictions, together with their improvement rates relative to the baseline ELDM calculations, are summarized in Table~\ref{tab:ml_results}.

\begin{table*}[htbp]
\caption{\label{tab:ml_results}Predictive performance of different machine-learning residual-correction models for $\log_{10}(T_{1/2})$. The improvement rates are calculated relative to the baseline ELDM results.}
\begin{ruledtabular}
\begin{tabular}{lcccc}
Model & RMSE & RMSE improvement (\%) & MAE & MAE improvement (\%) \\
\colrule
ELDM          & 0.567 & ---    & 0.417 & ---    \\
XGBoost-term1 & 0.478 & 15.700 & 0.354 & 15.234 \\
XGBoost-term2 & 0.475 & 16.236 & 0.347 & 16.775 \\
XGBoost-term3 & 0.379 & 33.187 & 0.282 & 32.339 \\
XGBoost-term4 & 0.386 & 31.853 & 0.283 & 32.132 \\
TabPFN-term1  & 0.460 & 18.949 & 0.330 & 20.977 \\
TabPFN-term2  & 0.454 & 19.905 & 0.323 & 22.595 \\
TabPFN-term3  & 0.348 & 38.597 & 0.248 & 40.461 \\
TabPFN-term4  & 0.353 & 37.710 & 0.253 & 39.258 \\
\end{tabular}
\end{ruledtabular}
\end{table*}

As shown in Table~\ref{tab:ml_results}, the introduction of machine-learning-based residual correction leads to a substantial improvement in the predictive accuracy of the ELDM baseline for $\alpha$-decay half-lives. Compared with the original ELDM calculations, all residual-correction models exhibit noticeable reductions in both RMSE and MAE, indicating that a considerable portion of the residual deviations can be effectively captured through the learning of microscopic nuclear-structure information.

From the overall comparison between different machine-learning frameworks, the TabPFN-based models generally demonstrate better predictive performance than the corresponding XGBoost models under the same feature combinations. In particular, the TabPFN-term3 model achieves the best overall performance, reducing the RMSE from 0.567 for the ELDM baseline to 0.348 and the MAE from 0.417 to 0.248. This result suggests that the probabilistic foundation-model framework possesses a strong capability for learning complex nonlinear residual structures from relatively limited nuclear datasets. It also indicates that the residual deviations of the ELDM calculations are not purely random fluctuations, but still contain systematic microscopic structure information that can be extracted through small-sample machine-learning approaches.

It can also be observed that the predictive performance improves with the progressive incorporation of physically motivated descriptors. In particular, the term3 feature combination consistently provides the best performance in both machine-learning frameworks, indicating that the simultaneous inclusion of deformation-related information, Geiger--Nuttall-related quantities, and orbital-angular-momentum descriptors plays an important role in the residual-correction process. Compared with term3, the slight performance degradation observed in term4 further suggests that removing the deformation-related descriptor weakens the ability of the model to capture deformation-associated residual structures. This confirms that nuclear deformation effects contribute non-negligibly to the residual deviations of the ELDM calculations.

These results suggest that the residual deviations between the ELDM predictions and experimental half-lives preserve important microscopic nuclear-structure information beyond the macroscopic barrier-penetration mechanism. Through the present physics-guided residual-correction framework, such missing microscopic effects can be partially compensated, thereby improving the predictive accuracy of $\alpha$-decay half-lives while maintaining the underlying physical interpretability of the ELDM-based framework.

\begin{figure*}[htbp]
\centering
\includegraphics[width=0.85\textwidth]{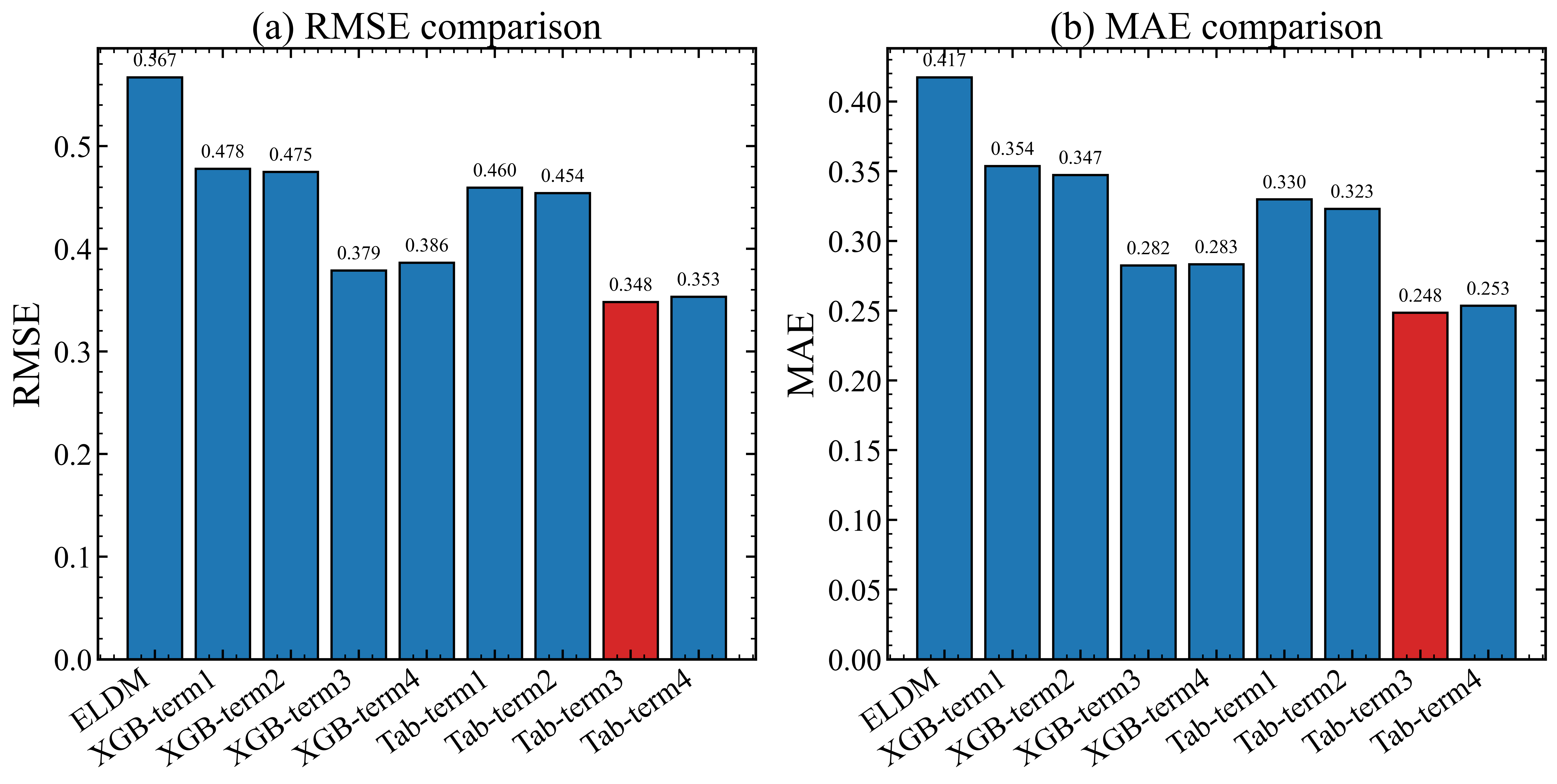}
\caption{\label{fig:performance_compare}
Comparison of the RMSE and MAE values for different machine-learning residual-correction models under various feature combinations.}
\end{figure*}

To provide a more intuitive comparison of the predictive performances of different models and their improvements relative to the baseline ELDM calculations, Fig.~\ref{fig:performance_compare} further presents the RMSE, MAE, and improvement-rate comparisons for all feature combinations. Overall, the predictive accuracy exhibits a clear improvement trend with the progressive incorporation of physically motivated descriptors, indicating that these physically meaningful features can effectively enhance the capability of machine-learning models in learning the residual structures of $\alpha$ decay. Among the different feature combinations, the term3 feature set consistently achieves the best performance in both the XGBoost and TabPFN frameworks, while the TabPFN-term3 model provides the overall highest predictive accuracy.

Meanwhile, the performance differences between the two machine-learning frameworks further reflect their distinct capabilities in learning complex residual structures. Compared with XGBoost, the TabPFN model demonstrates more stable predictive behavior under different feature combinations, especially after incorporating more complicated physical descriptors, where a more pronounced error-reduction trend can be observed. These results suggest that TabPFN possesses a stronger capability for modeling complex nonlinear residual relationships under limited-data conditions and can more effectively extract the microscopic nuclear-structure information hidden in the $\alpha$-decay residual deviations.

\begin{figure*}[ht]
\centering
\includegraphics[width=0.85\textwidth]{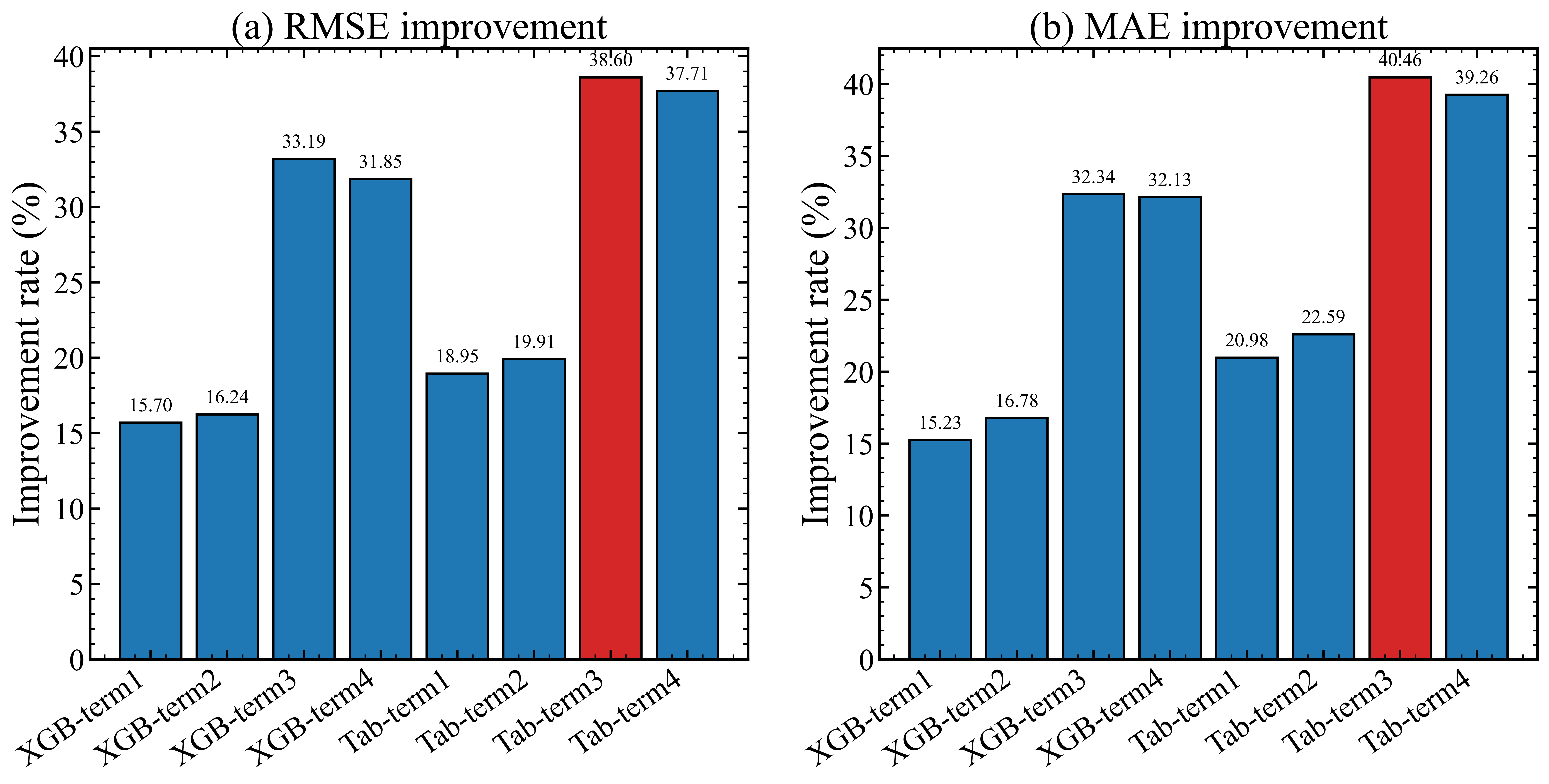}
\caption{\label{fig:improvement_compare}
Comparison of the improvement rates relative to the baseline ELDM calculations for different machine-learning residual-correction models under various feature combinations.}
\end{figure*}

\begin{figure*}[ht]
\centering
\includegraphics[width=0.55\textwidth]{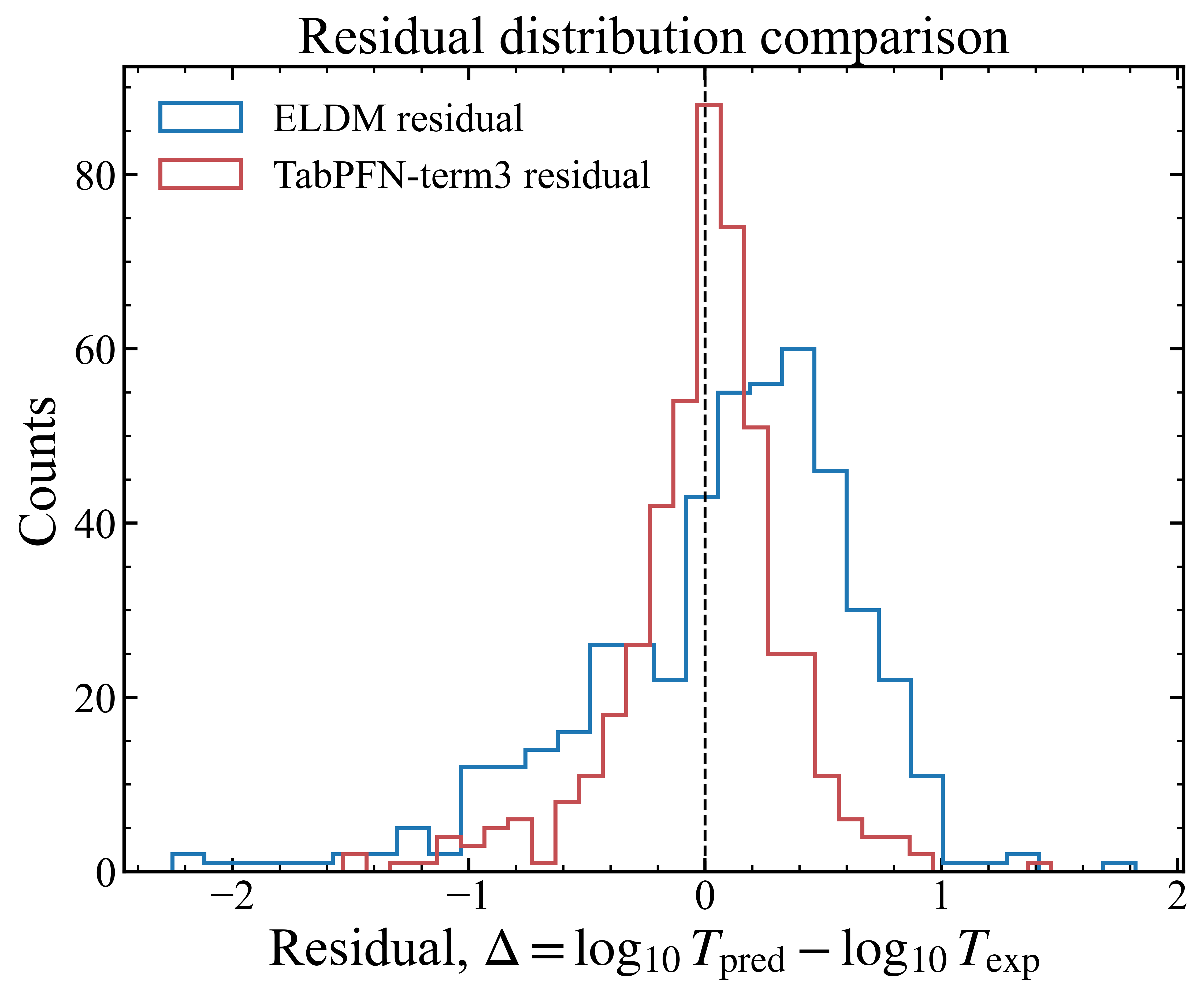}
\caption{\label{fig:residual_distribution}
Residual-distribution comparison between the baseline ELDM calculations and the TabPFN-term3 residual-correction results.}
\end{figure*}

To further investigate the influence of the machine-learning residual correction on the global error behavior, Fig.~\ref{fig:residual_distribution} presents the residual-distribution comparison between the original ELDM calculations and the TabPFN-term3 results. Compared with the baseline ELDM predictions, the residual distribution of the TabPFN-term3 model becomes significantly more concentrated around zero, while the distribution width in the large-residual region is substantially suppressed. This behavior indicates that the machine-learning residual correction can effectively reduce the systematic deviations between theoretical predictions and experimental data.

Meanwhile, the noticeable suppression of the long-tail residual region further demonstrates that the residual-correction framework is capable of mitigating extreme prediction deviations associated with complex microscopic nuclear-structure effects. As a result, the corrected predictions exhibit improved global stability and reliability for $\alpha$-decay half-life calculations. These results further suggest that the residual deviations in the original ELDM calculations still preserve important microscopic nuclear-structure information, which can be effectively extracted and compensated through the present physics-guided residual-correction framework.

\begin{figure*}[ht]
\centering
\includegraphics[width=0.85\textwidth]{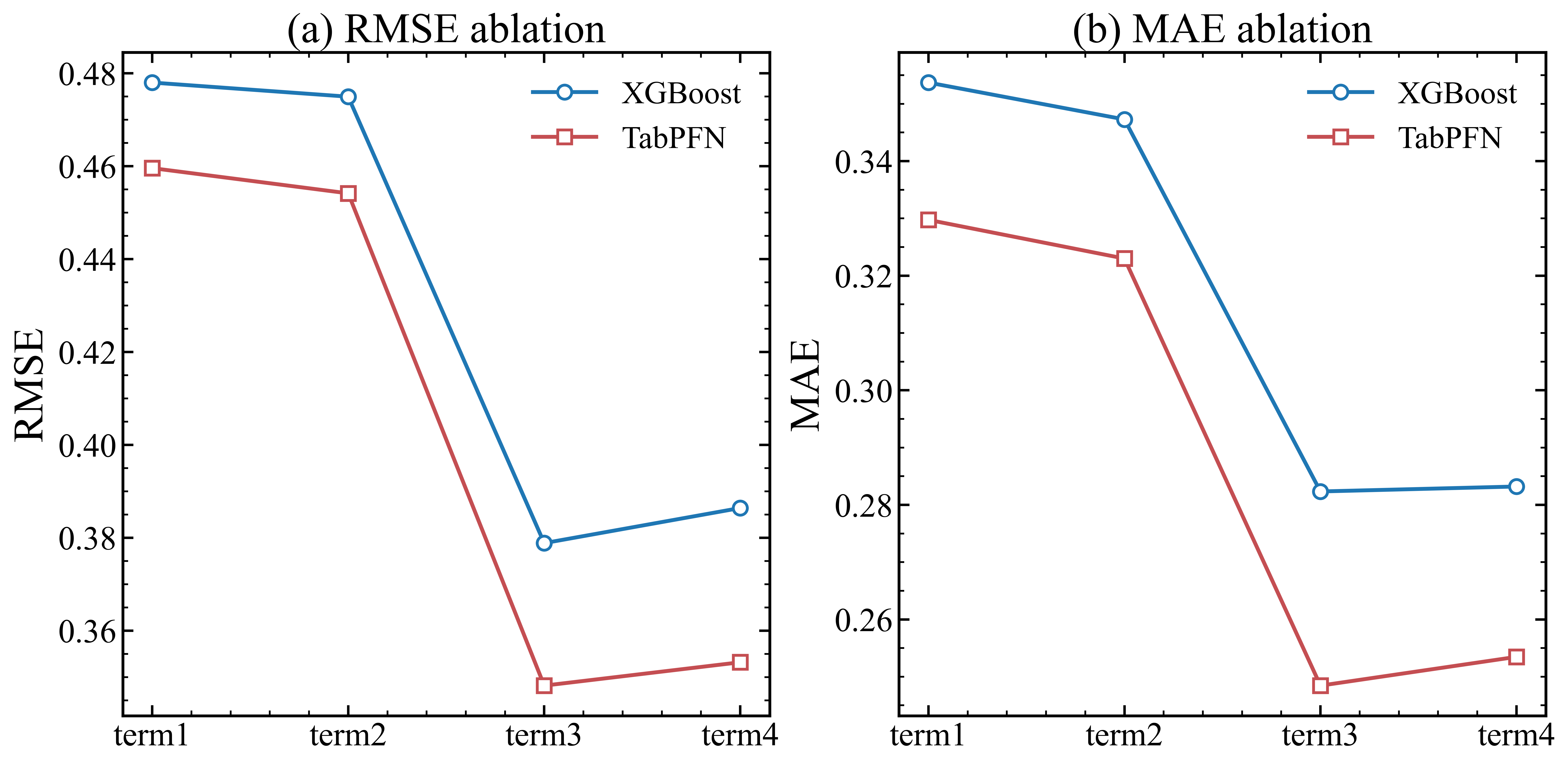}
\caption{\label{fig:feature_ablation}
RMSE and MAE variations under different residual-learning feature combinations for the XGBoost and TabPFN models.}
\end{figure*}

To further investigate the influence of different physically motivated descriptors on the residual-correction performance, Fig.~\ref{fig:feature_ablation} presents the RMSE and MAE variations under different feature combinations for both the XGBoost and TabPFN frameworks. It can be observed that the predictive performances of both machine-learning models exhibit a clear dependence on the adopted physical descriptors, indicating that different microscopic nuclear-structure information contributes differently to the residual-correction process.

For both machine-learning frameworks, the incorporation of additional physically motivated descriptors generally leads to improved predictive accuracy. In particular, the most noticeable performance improvement appears in the transition from term2 to term3, where the minimum orbital angular momentum descriptor $l_{\min}$ is further introduced. After incorporating the orbital-angular-momentum-related information, both RMSE and MAE are obviously reduced in the XGBoost and TabPFN calculations. This behavior indicates that the centrifugal hindrance effect associated with unfavored $\alpha$ decay plays an important role in the residual deviations of the ELDM calculations and can be effectively captured through the residual-correction framework.

Meanwhile, after removing the deformation-related descriptor in term4, a certain degree of performance degradation can still be observed compared with term3, especially in the TabPFN calculations. This behavior further demonstrates that nuclear deformation effects contribute non-negligibly to the residual structure of the ELDM predictions. In addition, the different sensitivities of XGBoost and TabPFN to deformation-related information suggest that different machine-learning frameworks may possess distinct capabilities in extracting microscopic nuclear-structure information hidden in the residual deviations.

Overall, the feature-ablation analysis further confirms that the residual deviations of the ELDM calculations still preserve important microscopic nuclear-structure information beyond the macroscopic barrier-penetration mechanism. Through the incorporation of physically motivated descriptors, the machine-learning models can effectively learn and compensate for these missing microscopic effects, thereby improving the predictive accuracy of $\alpha$-decay half-lives.

The above results further demonstrate that physically motivated descriptors play an essential role in the residual-correction process. In the present work, the ELDM baseline mainly describes the macroscopic barrier-penetration mechanism of $\alpha$ decay through an effective liquid-drop potential. In this framework, the prescission region is characterized by the liquid-drop energy variation of the dinuclear configuration, while the external region is mainly governed by the Coulomb barrier and, when necessary, the centrifugal correction. Although the ELDM can reasonably reproduce the global tunneling behavior of $\alpha$ decay, several microscopic nuclear-structure effects are still not fully incorporated in the baseline calculations. Consequently, systematic residual deviations may appear for heavy and superheavy nuclei with more complicated structural characteristics.

By progressively incorporating deformation-related information, Geiger--Nuttall-related descriptors, and orbital-angular-momentum information into the residual-correction framework, the machine-learning models become capable of learning part of the microscopic correction effects absent in the conventional ELDM calculations, thereby effectively reducing the systematic deviations between theoretical predictions and experimental data. In particular, the term3 feature combination, which simultaneously contains multiple physically motivated descriptors, produces the most significant improvement in predictive accuracy, especially in the TabPFN calculations. This behavior suggests that different microscopic nuclear-structure effects may exhibit coupled nonlinear characteristics in the residual structure of $\alpha$-decay half-life predictions, and such complex correlations can be effectively captured through small-sample residual-learning approaches.

Overall, the present physics-guided residual-correction strategy can effectively compensate for the missing microscopic nuclear-structure effects while preserving the macroscopic physical constraints and interpretability of the ELDM framework. Such a hybrid framework provides a feasible and physically interpretable approach for improving high-precision $\alpha$-decay half-life predictions in heavy and superheavy nuclei.

\section{Conclusion}

In the present work, a physics-guided residual-correction framework combining the ELDM with machine-learning models was developed for the high-precision prediction of $\alpha$-decay half-lives in heavy and superheavy nuclei. The ELDM calculations were first employed as the macroscopic baseline description of the barrier-penetration process, while XGBoost and TabPFN models were subsequently introduced to learn the residual deviations between theoretical predictions and experimental data.

Several groups of physically motivated descriptors, including deformation-related quantities, Geiger--Nuttall-related features, and orbital-angular-momentum descriptors, were systematically constructed for the residual-correction calculations. The results show that all machine-learning residual-correction models can significantly improve the predictive accuracy of the original ELDM baseline. Among them, the TabPFN-term3 model achieves the best predictive performance, with RMSE and MAE values reduced to 0.348 and 0.248, corresponding to improvements of 38.60% and 40.46%, respectively, relative to the baseline ELDM calculations.

Residual-distribution analysis further demonstrates that the machine-learning correction can effectively suppress large-deviation regions and improve the overall stability and reliability of the predictions. In addition, feature-ablation analysis indicates that the incorporation of physically motivated descriptors plays an important role in the residual-correction process, suggesting that the microscopic nuclear-structure effects not fully incorporated in the baseline ELDM calculations can be partially compensated through the present residual-learning framework.

Overall, the present study demonstrates that the combination of a physically interpretable macroscopic model and small-sample machine-learning approaches provides an effective and physically meaningful strategy for improving the prediction accuracy of $\alpha$-decay half-lives. The proposed physics-guided residual-correction framework may also provide useful guidance for future investigations of other nuclear-decay processes and related nuclear-structure problems.

\section{Acknowledgements}

This work is supported by Yunnan Fundamental Research Projects (No. 202501AT070067, 202401AU070074 and 202501CF070189), Yunnan Provincial Xing Dian Talent Support Program (Young Talents Special Program, (Young Talents Special Program, No. XDYC-QNRC-2023-0162), Kunming University Talent Introduction Research Project (No. YJL24019), Yunnan Provincial Department of Education Scientific Research Fund Project (No. 2025Y1055 and 2025Y1042), the Special Basic Cooperative Research Programs of Yunnan ProvincialUndergraduate Universities’ Association (NO. 202101BA070001-144), the Program for Frontier Research Team of Kunming University 2023, National Natural Science Foundation of China (No. 12063006), National College Student Innovation and Entrepreneurship Training Program (No. 202511393011, 202511393012, and 202511393015), Yunnan Province College Student Innovation and Entrepreneurship Training Program (No. S202511393003, S202511393043, and S202511393044), and Xing Dingyu Academician Workstation of Yunnan Province (No. 202605AF350035).

% The \nocite command causes all entries in a bibliography to be printed out
% whether or not they are actually referenced in the text. This is appropriate
% for the sample file to show the different styles of references, but authors
% most likely will not want to use it.
\nocite{*}

\bibliography{apssamp}% Produces the bibliography via BibTeX.

\end{document}